\def\ut#1{\mathop{\vtop{\ialign{##\crcr
     $\hfil\displaystyle{#1}\hfil$\crcr\noalign
     {\kern1pt\nointerlineskip}\hbox{$\hfil\sim\hfil$}\crcr
     \noalign{\kern1pt}}}}}

\def\undersymbol#1#2{\mathop{\vtop{\ialign{##\crcr
     $\hfil\displaystyle{#2}\hfil$\crcr\noalign
     {\kern1pt\nointerlineskip}\hbox{$\hfil#1\hfil$}\crcr
     \noalign{\kern1pt}}}}}
\def\arcsec{^{\prime\prime}}

\def\hour{^{\rm h}}
\def\minute{^{\rm m}}

\documentclass{ws-procs961x669}            
\begin{document}
\title{NEVER  BET AGAINST EINSTEIN\,\footnote{This paper is based on a ``popular'' talk delivered at the Fourth Punjab University International Conference on Gravitation and Cosmology (4th PUICGC) on November 22, 2021, dedicated to Prof. Asghar Qadir on the occasion of his 75$^{\rm th}$ birthday.}}

\author{Francesco De Paolis}

\address{Department of Mathematics and Physics 
``E. De Giorgi'', University of Salento, Via Arnesano, 73100 Lecce, Italy \\ and \\
INFN, Sezione di Lecce, Via Arnesano, 73100 Lecce, Italy\\ and \\
INAF, Sezione di Lecce, Via Arnesano, 73100 Lecce, Italy\\
francesco.depaolis@le.infn.it}

\begin{abstract}
Among the geniuses of mankind, Einstein was probably one of those who made more erroneous claims, and often changed his opinion during the years on important scientific subjects. However, it is important to bear in mind that his mistakes were always due to a misinterpretation of the results obtained from his equations, since he was often biased by his own view of the universe. Einstein’s equations have  always given the correct answer and, till now, more than a century after its development, there is no evidence that it may  not be correct.
\end{abstract}


\bodymatter

\section{Introduction}	
At the end of  1915 Albert Einstein completed his theory of General Relativity (GR), which appeared in printed form between the end of that year and the beginning of 1916. \cite{Einstein1915,GR} 
At that point, the magic reach of this theory appeared clearly, and a  succession of predictions, most of which might have looked as bold leaps arised rapidly.
As a matter of fact, indeed,  evenl at the end of the seventies of the last century many scientists did not take seriously many of those predictions, such as the existence of gravitational waves. Let us look at some of Einstein's  predictions.\footnote{For a very recent book on GR with emphasis towards an historical approach, the reader is referred to Ref.~ \refcite{Qadir2020}.}

The starting point of Einstein's predictions was that of considering the effect of the curvature induced by the solar gravitational field on the motion of the planets of the Solar System: the motion predicted by Kepler's laws was recovered also in GR, but with some deviations, in particular for the planets closer to the Sun. As far as Mercury is concerned, in fact, Einstein found a precession of the perihelion of $0.43\arcsec$ per century, in perfect agreement with the observed shift, which was not possible to account for by Newtonian dynamics. Actually, this was not a truly novel prediction since the anomaly of Mercury's perihelion precession was known, even if no explanation had been found for it. The really new prediction of GR are discussed in the following Sections.

\section{Gravitational lensing}

Einstein considered the deflection of light rays passing close to the Sun disk. Actually, already before completing GR,  Einstein correctly understood that a massive body (for example the Sun) may deflect  light rays passing close to the body's surface. Already in 1911 he had calculated the effect on the basis of Newtonian mechanics \cite{Einstein1911} and found a value of about $0.87\arcsec$ for a star close to the Sun's limb.\footnote{We note here that the first qualitative discussion of the light deflection within Newtonian mechanics, by Newton himself,  goes back to 1704, while the first calculation of the deflection angle was made by Soldner in 1801 and published in 1804. \cite{Soldner}} Einstein then wrote to Hale, the renowned astronomer, inquiring whether it was possible to measure a deflection angle of about $0.87\arcsec$ for a star towards the Sun's limb. Of course, the answer was negative, but Einstein did not give up, and when, in 1916, he made the calculation again using GR he found the right value,  that is  about $1.75\arcsec$. That result was resoundingly confirmed during the Solar eclipse of May 29, 1919 \cite{Dyson}.

The story goes on and in 1924 Chwolson, in a paper almost unnoticed, \cite{Chwolson1924} considered the particular case when  source,  lens and observer are aligned and noticed the possibility of observing a luminous ring when a far source undergoes the lensing effect by a massive body in between. In 1936 Einstein published a paper  describing the gravitational lensing effect of one star on another (this effect is called at present gravitational microlensing), the formation of the luminous ring, today called the Einstein ring, and giving the expression for the source amplification.\cite{Einstein1936} However, Einstein considered this effect exceedingly curious and useless since, in his opinion, there was no hope to actually observe it: one had to observe a star for millions of years to possibly see a microlensing event and, moreover, the source images produced by the lens appear too close to the source star to have a chance to be detectable. It is interesting to note, however, that both predictions are wrong, but it took several decades to observationally confirm that microlensing events can be detectable and that the images can be really observed and separated from the source star brightness. It is true that both Einstein's predictions were wrong, but only because he underestimated the technological progress and did not foresee the motivations that today induce astronomers to widely use the gravitational lensing phenomenon. 

However, at variance with Einstein, Zwicky promptly understood that galaxies were gravitational lenses more powerful than stars and might give rise to images with a detectable angular separation. In two papers published in 1937, \cite{Zwicky1937a,Zwicky1937b} Zwicky noticed that the observation of galaxy lensing, in addition to giving a further proof of GR, might have allowed observing sources otherwise invisible, thanks to the light gravitational amplification, thereby obtaining a more direct and accurate estimate of the lens galaxy dynamical mass. He also found that the probability to observe lensed galaxies was much greater than that of star on star. This shows the foresight of this eclectic scientist, since the first strong lensing event was discovered only in 1979: the double quasar QSO 0957+561 a/b, \cite{qso0957+561} shortly followed by the observation of tens of other gravitational lenses, Einstein rings and gravitational arcs. 

All of that plays today an extremely relevant role for the comprehension of the formation and evolution of the structures in the universe, for the measure of the parameters of the so-called cosmological standard model, and also for testing  GR and alternative gravity theories (see, e.g., Ref. \refcite{Collett2018} and references therein).

Since the last part of the twentieth century, gravitational microlensing has become a way to constrain the abundance of the so-called MACHOs (Massive Astrophysical Compact Objects) in the halo of our and nearby galaxies and to map the star distribution towards the Galactic Bulge, thorough various Projects such as MACHO (MAssive Compact Halo Objects), \cite{Alcock1993} EROS (Expérience pour la Recherche d’Objets Sombres), \cite{Aubourg1993} OGLE (Optical Gravitational Lensing Experiment), \cite{Udalski1993} MOA (Microlensing Observations in Astrophysics), \cite{Bond2001} DUO (Disk Unseen Objects), \cite{Alard1995} WISE (Wide-field Infrared
Survey Explorer), \cite{Shvartzvald2012} KMTNet (Korea Microlensing
Telescope Network), \cite{Kim2018} MEGA (Microlensing Exploration of the Galaxy and Andromeda), \cite{deJong2004} AGAPE (Andromeda Galaxy Amplified Pixel Experiment), \cite{Ansari1997}  WeCAPP (Wendelstein Calar Alto Pixellensing Project), \cite{Lee2012} PLAN (Pixel Lensing towards Andromeda), \cite{CalchiNovati2014} to cite only some of them.

 In the last decades, moreover, gravitational microlensing has become a standard technique adopted nowadays to discover new exoplanets, being complementary  to the other techniques used by exoplanet hunters, which  are radial velocity, direct imaging, transit and astrometry (for a comprehensive review on extrasolar planets see Ref.~\refcite{Perryman}), to mention the most relevant ones.

Until now, more than about 160 exoplanets (eight of which are multiple planet systems) have  been discovered by gravitational microlensing.\footnote{The reader may look at  the website http://exoplanet.eu/, which is maintained and continuously updated by the {\it Exoplanet TEAM}.} 
In the case of binary microlensing events, such as those produced by a lensing stars with a planet around, the number and the position of the images  differ from those of the single lens case and the astrometric signal
trajectory and the deviation varies depending on  the
binary system parameters. \cite{Han1999,Bozza2001,Sajadian2015}
 Identifying lens binarity is generally not so difficult, in particular in the case of events characterized by caustic crossing whose light curve may show  strong deviations with respect to the single-lens Paczyński light curve. However, light curves with minor deviations from a Paczyński-like shape do not allow one to identify the source binarity. An important consequence of gravitational microlensing is the shift of the position of the  image centroid with respect to the source star location. This effect gives rise to the  so-called astrometric microlensing signal. When the astrometric signal is considered, the presence of a binary source manifests itself with a path that substantially differs with respect to that expected for single source events. As a matter of fact, considering the  astrometric signatures of binary sources  and taking into account their orbital motion and the parallax effect due to the Earth’s motion, turns out to be very important   in the analysis of astrometric data in order to correctly estimate the lens-event parameters. For further details on this issue we refer the interested reader to the analysis presented in Ref. ~\refcite{Nucita2016}.
It is also worth mentioning that generally a static binary lens is considered in the fittiing procedures to the observed light curves. However, it is often important to treat the binary-lens motion in a realistic way. It can be shown that an accurate timing analysis\,\footnote{Timing analysis is a powerful technique widely used in astrophysics to determine periodic features of different kind of signals. The application of the Lomb-Scargle \cite{Lomb} and/or Schwarzenberg-Czerny \cite{Czerny} periodograms to microlensing light curves may allow obtaining in an independent way the orbital period and/or the spin period of the lensed star by observing the signatures induced by the stellar spots on the star surface. \cite{Nucita2014,Giordano2015}} of the residuals (calculated with respect to the best-fitting Paczyński model) is usually sufficient to infer the orbital period of the binary lens and the estimated period  may be used to further constrain the orbital parameters obtained by the best-fitting procedure, which often gives degenerate solutions. \cite{Giordano2017}

Since gravity is a natural cosmic lens, it can magnify consistently far away stars, and even stars in other galaxies. The most impressive of such cases is represented by MACS J1149+2223 Lensed Star 1, a star in a far galaxy at redshift $z \simeq  1.49$ magnified by more than $2\times 10^3$ times by the gravitional field of a foreground, massive cluster of galaxies.\,\cite{Kelly2018} It goes without saying that such star could never be visible in the absence of the microlensing magnification and that the observations of such phenomena may shed light on the dark matter question, extremely relevant in astrophysics and cosmology.

Gravitational microlensing is so powerful a technique that can allow detecting exoplanets not only in the Milky Way but also in  nearby galaxies such as M31, as in the case of the evsnt PA-99-N2 discovered by the POINT-AGAPE Collaboration\,\cite{An2004} and re-analyzed in a novel way some years later. \cite{Ingrosso2009} The result of a Monte Carlo based analysis of the data acquired in 1999 have shown that the anomaly observed in the light curve with respect to the Paczynski behaviour can be accounted for by the presence of an extrasolar planet with mass about 6.34 times  that of Jupiter around a  microlensing  star of about $0.5\,M_{\odot}$, most likely situated in the M31 galaxy (see also Ref.~\refcite{Ingrosso2011}). To our knowledge, there exists only  another extragalactic planet candidate, recently discovered in the M51 galaxy by analyzing the data showing a transit-like lightcurve of an X-ray source (called M51-Uls-1) detected  by XMM-Newton.\cite{DiStefano2021} Since the estimated orbital period of this exoplanet is of about 70 years, further analysis is necessary in order to confirm this discovery. 

Note also that gravitational microlensing represents practically the only way to discover free-floating planets or even small-mass primordial black holes (PBHs)  either in our or nearby galaxies. \cite{Sumi,Hamolli2015,Hamolli2016,Niikura2019} Indeed, free-floating planets, or rogue planets, are expected to form in planetary systems and then be ejected during the early stages of the planetary system formation. Their mass being,  according to current planetary formation theories, in the range between about 0.1 and 1 Earth masses, they are extremely difficult to be detected. Microlensing events due to these light objects are expected to have extremely short timescales. The shortest duration of such events is OGLE-2016-BLG-1928 with Einstein time about 41.5 minutes and estimated mass between that of Mars and the Earth mass, with the last possibility favoured by considering the recent proper motion measurement of the source star by the Gaia space-based telescope. \cite{Mroz2020} 

Among the exoplanets detected by gravitational microlensing we would like to mention the particularly important  event recently discovered towards the Taurus constellation \cite{Nucita2018}, lying in the opposite direction with respect to the Galactic Bulge where most microlensing events are found (due to the larger stellar optical depth).
The importance of this event relies on the fact that it is the closest microlensing event ever found and, due to this region, it is the only event \footnote{We mantion that very recently the same procedure has been applied to the case of the Gaia19bld event, \cite{Rybicki2021} for which the interferometric measurements are presented in Ref. ~\refcite{Cassan2021}.} for which the two microlensing images have been separated.\cite{Dong2019} This result has been made possible due to the VLTI Gravity instrument which combines interferometrically 4 telescopes of 8.2 meters each, that is equivalent to a single mirror of 130 meters size.
We emphasize that this result also contradicts Einstein expectation of 1936 that microlensing images cannot be resolved since they are too close to the source star.

A novel tool of gravitational microlensing is the so-called  {\it  astrometric microlensing}. It is well known that gravitational redshift is a consequence of GR and it was confirmed for the first time by Adams who measured, already in 1925, the light redshift from the surface of Sirius B. \cite{Adams1925} It was then confirmed also in laboratory experiments.\cite{Pound} Concerning Adams's important discovery, Eddington later wrote that ``Prof. Adams has killed two birds with one stone: he has carried
out a new test of Einstein’s general theory of relativity and he has confirmed our
suspicion that matter 2000 times denser than platinum is not only possible, but is
actually present in the universe'', thereby confirming the existence and the physical properties of white dwarfs. \cite{Eddington} More recently, the passage of another white dwarf (Wd Stein 2051B) in front of a background star at a distance of about 1.6 kpc has been used for the first time to weight the white dwarf by measuring with the Hubble Space Telescope (HST) the position changes of the star caused by the gravitational deflection. The resulting white dwarf mass was estimated to be $\simeq 0.675\,M_{\odot}$ with an uncertainlty of about $8\%$. \cite{Sahu2017} This important effect had not been predicted by Einstein but it is written in the equations of GR and is nowadays called  astrometric microlensing. It will certainly play growing importance in astrophysics in the near future. \cite{Safizadeh1999,Dominik2000,Sajadian2015,Nucita2016,Nucita2017}

Lensing and  microlensing of quasars are also important tools in astrophysics and cosmology since they  may allow studying in detail both the quasars and the mass distribution of the lensing galaxy. Particularly relevant in this respect is the 2020 discovery of an optical flare by the Caltech’s robotic Zwicky Transient Facility (ZTF),  \cite{Graham2020} towards the active galactic nucleus J1249+3449. This event was
claimed to be the electromagnetic counterpart associated with a binary
black hole merger detected by the LIGO (Laser Interferometer Gravitational Wave Observatory) and Virgo Collaboration about 47 days before the optical flare. A subsequent analysis of the acquired data, however, shows that data are more consistent with the possibility that the optical flare is a quasar microlensing event
 probably due to a $\simeq 0.1\,M_{\odot}$ massive object amplifying the light from the active galactic nucleus. The interested reader is referred to Ref.~\refcite{DePaolis2020}  for further details. We also remark that the study of this kind of transient events towards far away quasars is particularly relevant in the context of the next-coming large surveys of the
sky, such as that provided by the Vera C. Rubin Observatory with
the Legacy Survey of Space and Time (previously referred to as LSST) survey, which will
soon produce a survey of the whole Southern sky every few days detecting at least $10^7$ AGNs and quasars up to a limiting magnitude of $\simeq 24.1$ and redshift $z \leq 2.1$. \cite{Abell2009}

Gravitational microlensing constitutes only one of the several scales in which the gravitational lensing phenomenon may manifest itself, and that may lead obtaining valuable information about a  variety of astronomical issues ranging from the star distribution in the Milky Way, to the study of stellar atmospheres, the discovery of exoplanets in the Milky Way and also in nearby galaxies, the study of far away galaxies, galaxy clusters and black holes (BHs). For a more detailed account of all these issuees the reader is addressed to Ref.~\refcite{DePaolisUniverse}.
 
Just to give an example, the weak lensing effect, that is the deformation of far away galaxies  due to the mass distribution in between, is nowadays an extremely important technique for studying the distribution of  dark matter and dark energy in cosmology. Two forthcoming space-based missions devoted almost completely to weak lensing are the NASA Nancy Grace Roman WFIRST telescope and the ESA Euclid mission. We also note that these space-based telescopes will be also used for microlensing searches (see, e.g., Refs ~\refcite{Yee2013,Penny2013,Bachelet2019}, and references therein).

All lensing cases considered until now are weak field effects and indeed they always involve small angle deflections of the light rays from far away sources. However, if the lens is a BH, photons can be deflected by large angles and, if they  pass very close to the BH. These light rays can even go backwards, once, twice or more times, giving rise, in the case of observer, lens and source perfectly aligned, to a series of concentric rings with decreasing brightness from the outermost to the innermost ring.  This phenomenon gives rise to the so called retrolensing events. In this respect, in the last scientific paper  by John Archibald
Wheeler, written in collaboration with Daniel Holz, \cite{HolzWheeler2002} it was proposed to perform a survey of the full sky searching for concentric rings as a way to discover BHs around the Solar System (for the gravitational lensing of light rays in the strong field regime see, e.g., Refs. ~\refcite{Bozza2001L,Bozza2002}). However, it is not difficult to show that it is very hard to discover
nearby BHs in such a  way since the
maximum distance $D_L$ at which the retro images due to a BH with mass $M_{BH}$ can be seen by an instrument with limiting magnitude $\bar{m}$ turns out to be
\begin{equation}
D_L= 0.02\, {\rm pc}\times \left\{10^{(\bar{m}-30)/2.5}\left(\frac{M_{BH}} {10\,M_{\odot}}\right)^2 \right\}^{1/3}
\end{equation}
for an assumed image baseline of magnitude $m\simeq 30$.\cite{DePaolis2003}  Therefore, only a BH heavier than about $10\,M_{\odot}$ lying within about $10^{-2}$ pc from Earth might
be revealed in this way with the present instruments, and we already know that no such a massive BH can exist so close to our Solar System! 

A more clever idea is that of considering the supermassive BH Sgr A$^*$ lying at the Galactic center. 
Observations show that this BH has a mass of about $(4.6\pm 0.7) \times 10^6\,M_{\odot}$ and is surrounded by  many stars orbiting around it with very high speeds. \footnote{The reader is referred to the webpage https://www.astro.ucla.edu/~ghezgroup/gc/ for further details.} In fact, it is the study of the motion of these stars that allows to infer the physical parameters of the  Sgr A$^*$ BH and Andrea Ghez and Reinhard Genzel won the 2020 Nobel prize in Physics for their (with their teams) study of the compact object at the Galactic center.\cite{Schodel2002,Ghez2003,Eisenhauer2005,Ghez2005,Gillessen2017}
The most recent analysis of the simultaneous detection, within the diffraction limit of the four-telescope interferometric beam combiner GRAVITY/VLT, of the four stars S2, S29, S38 and S55 orbiting the central BH has allowed obtaining a high-precision determination of the gravitational potential around Sgr A$^*$. The obtained results are in excellent agreement with the  GR prediction of stellar orbits around a single central point with mass about $M = 4.30 \times 10^6 \,M_{\odot}$, with a precision of about $\pm 0.25\%$. \cite{Gravity2021} 

In addition to Ghez and Genzel, the 2020 Nobel prize in Physics was awarded also to Roger Penrose, for the “discovery that BH formation is a robust prediction of the general theory of relativity”. If ever needed, this prize certifies the existence of BHs in the universe. Note also that many scientists and even Einstein himself did not believe in the existence of BHs and in 1939 he had written that “It is a clear understanding as to why the Schwarzschild singularities do not exist in physical reality.” \cite{Einstein1939} Einstein's conclusion in this respect was based on the analysis within GR of the contraction of a spherical cluster, whose component velocity reach the light speed when the radius of the cluster turns out to be about 1.5 times the Schwarzschild radius. Therefore, their formation seemed forbidden. It is however worth mentioning that BHs cannot form in this way and their formation always has to involve violent and non-equilibrium processes. Indeed, it is now a widely accepted fact that BHs are a robust prediction of GR.

The Sgr A$^*$ BH may also act, in principle,  as a strong gravitational lens on the light rays from the stars orbiting around it, as for example the S2 (also referred to as SO-2) star, which makes an orbit around Sgr A$^*$ in about 15.6 years. It was found that the retrolensing image by the Sgr A$^*$ BH of the  S2 star may have magnitude in the range $33-37$ in the K band (which is centered at wavelength $\lambda\simeq 2.2\,\mu$m) and could be eventually detectable by the next generation instruments such as the JWST.\cite{DePaolis2003} As a matter of fact, indeed, the relatively vicinity of S2 to the central massive BH may offer a unique laboratory to test the formation of retro-lensing images, and it has been recently announced that one of the key projects of JWST is to probe the  BH at the Galactic center.
The analysis of the shape of the retrolensed images can allow constraining the BH parmeters, i.e.  its spin, \cite{DePaolis2004,Bozza2004}
and its electric charge, \cite{Zakharov2005,Abdujabbarov2017,Tsukamoto2022,Babar2021} by measuring the angular size and the shape of the retrolensed image. Application of the retrolensing phenomenon to even  more exotic objects, such has wormholes and naked singularities have been proposed and discussed in the literature (see, e.g.,  Refs. ~\refcite{Tsukamoto2017a,Tsukamoto2017b}).
The retrolensing images define, in practice, the inner boundary of the BH shadow (see also Ref. ~\refcite{Zak2005}), which was first calculated in 2000  by Falcke, Melia and Agol, \cite{Falcke2000} and spectacularly observed by the EHT Collaboration in the case of the supermassive BH, with mass about $(6.5 \pm 0.7) \times 10^9\,M_{\odot}$, at the center of the M87 galaxy. \cite{EHT}
This discovery, together with the gravitational wave (GW) detection  (see Section 3), marks the beginning of a new era in BH astrophysics allowing for new tests of gravity in strong field regimes. Observations are in perfect agreement with the GR predictions and many alternative  gravity theories have been ruled out by the EHT measurements (see, e.g., Ref.~\refcite{Psaltis2020} and references therein).

Moreover, while gravitational retrolensing produced by  a Schwarzschild BH always conserves the ``color'' of the source star, in the case of a Kerr BH it does not hold anymore.\footnote{We note that gravitational lensing is in general an achromatic effect since the deflection angle of a light ray does not depend on its wavelength. This is not more true for light rays passing close enough to a Kerr BH.} Indeed, it can be shown that  by measuring the color difference of one side of a retrolensed image by a Kerr BH with respect to the other side one could infer, in an independent way, the BH spin. \cite{DePaolis2011}

We also notice that a classical method for estimating the physical parameters (in particular mass and
angular momentum) of BHs is that of measuring the periastron or apoastron shift of surrounding test masses, analogously to Mercury's perihelion precession  which was  one of the first tests of GR. In particular, in the case of the stars orbiting around the Sgr A$^*$ BH, one can show that the
amount of the apoastron shift substantially depends not only on the BH mass but also on the distribution of both the stars
and the dark matter arond the  Galactc center, making it practically impossible to estimate the BH parameters through this method. 
In fact, as discussed in Ref. ~\refcite{Nucita2007}, the difference between the periastron shifts for the Schwarzschild and a maximally rotating Kerr BH is at most of about $10\,\mu$arcsec in the case of the  S2 star. In order to make these measurements
with the required accuracy, one would need to measure the  S-star orbits with a precision of at least $10\,\mu$arcsec. Such a precision is not far from that reachable in the near future.
However, the effect of the stellar cluster distribution around the central BH gives, for almost all the possible configurations,  a much larger effect on the periastron shift with respect to that of the central BH, thereby making very hard to measure the central BH physical parameters by this technique. \cite{Nucita2007} 
An analogous effect is also induced by the possible presence of a cluster of dark matter particles about the central BH. \cite{Nucita2007,DePaolis2011}
However, we mention that the orbits of S-stars may be used to test the form of the gravitational field toward the Galactic center and derive hints on the presence of a BH or a self-gravitating dark matter distribution, as suggested for example in Ref. ~\refcite{RAR2015}.
Indeed, the fact that the S-stars trajectories have roughly elliptical shape and the foci of the orbits are approximately coincident with the position of the Galactic center indicate that the central potential should be Newtonian with a BH in the center, as found in Ref.~\refcite{Zakharov2021}.

\section{Gravitational waves}
GWs are spacetime distortions occurring when massive objects like BHs collide or merge. The GWs produced squeeze and stretch space as they pass, and these effects are now detectable by the interferometers such as LIGO and Virgo. This led to the 2017
Nobel prize in physics to Barry Barish, Kip Thorne and Ray Weiss. 

GWs are another effect predicted by GR. The existence and propagation properties of GWs were calculated for the first time by Einstein in 1916. \cite{Einstein1916} He adopted the weak field linearized approximation and made simplifying assumptions regarding the gravitational field and obtained the approximate solutions of plane GWs travelling at the speed of light by introducing the harmonic coordinate condition into the field equations. However, later on, in 1936, together with Rosen,  he tried to solve the  non-linear field equations and find exact plane GWs. In fact, a fundamental problem which bothered Einstein since 1916 was whether the fully nonlinear field equations admitted solutions that can be interpreted as GWs. In case of an affermative answer to that question than, of course, far from the GW sources it results to be entirely reasonable to use  the linearized theory while in the other case it makes no sense to  spend time and efforts  to try to detect GWs since the solutions of the linearized theory are only artifacts of the linearization process.
In the full nonlinear GRin vacuum, Einstein and Rosen were indeed able to finds  a  solution representing a plane polarized GW, but it was necessary to introduce some singularities into the components of the metric describing the GWs. Due to the presence of these singularities, Einstein convinced himself (erroneously) that no
exact plane GW solutions to the field equations  exist. Einstein and Rosen then sent a paper with title ``Do Gravitational Waves Exist?'' to  Physical Review that, luckily enough, rejected it, provoking a furious reaction by Einstein, who never published another paper in that journal.\footnote{To be precise, in 1952 Einstein published again in the Physical Review,  but it was only  a comment to a paper by C.P. Johnson.\cite{Einstein1953}}  Still at the end of 1936, in a talk in Princeton on the 
“Nonexistence of gravitational waves'', Einstein's conclusion was that “if you ask me whether there are gravitational waves
or not, I must answer that I don’t know. But it is a highly interesting problem.”

As a matter of fact, in the equations of GR the answer was clear and GWs indeed exist. In fact, shortly after the accident with Physical Review, Howard Robertson (who was the referee of the rejected paper) met Einstein's collaborator Leopold Infeld and clarified the mistake in that paper. The non-linearized
approximation does indeed lead to plane transverse GWs. However, one cannot construct a single coordinate system to describe plane GWs without introducing a singularity somewhere in the spacetime. Today it is understood that such a singularity is only an apparent and not a real singularity. It is, indeed, a coordinate singularity, and while nowadays any student knows the difference between coordinate and physical singularities, in the 1930's no mathematical formalism for distinguishing the two were available and Einstein and Rosen had not clear in mind in 1936  that there was no reason to try to cover the whole spacetime  with a single coordinate system. By the way, it was  Robertson himself who suggested the ``trick'' of trasforming the Einstein-Rosen metric from space-time coordinates, suitable for representing plane GWs, to cylindrical coordinates. The singularity could then be located at
the origin of the cylindrical axis, where one would expect to find the source of the cylindrical waves. In this way the singularity can be regarded as describing a material source. The solution obtained can be considered describing cylindrical GWs rather than plane GWs.  Finally, Einstein had convinced himself that GWs were real and sent the corrected paper for publication. \cite{Einstein1937} Note that Einstein could have found the above escape to cylindrical waves months before, simply by reading the Physical Review referee's report, which he had dismissed so hastily and at the end of the paper \cite{Einstein1937} he acknowledged Professor Robertson for his friendly assistance in the clarification of the original error. For some historical reviews of this issue see Refs.~\refcite{Kennefick2005,Hill2017}

 Exactly one century after the first Einstein's prediction of GWs, they were detected by the LIGO/Virgo Collaboration. The first event, GW150914, detected on September 14 2015, certifies that Einstein was indeed right. \cite{Abbott2016} 
 This detection, together with the many others by the LIGO/Virgo and, since 2020, by the LIGO/Virgo/KAGRA Collaboration (KAGRA is the acronym of Kamioka Gravitational Wave Detector), have disclosed a new way to probe the universe, in addition to the traditional astronomical observations in the different bands of the electromagnetic spectrum, neutrinos and cosmic rays. GW detection has given rise to the so-called ``multimessenger astronomy'' and has hallowed, for the first time, not only to ``see'' black hole and neutron star binaries in the last stage of their coalescing process, but also the final black hole formation from the merging phenomenon. 
 
 A particularly important event was, indeed, GW170817, the first event produced by the coalescence of two neutron stars. The system is somehow similar, even if at a much later evolution time, to the binary neutron star system PSR B1913+16, \cite{Hulse1974} which has led to the first indirect detection of GWs \footnote{We mention that, very recently, the data analysis of the  double pulsar PSR J0737-3039, discovered in 2003, \cite{Burgay2003} consisting of two radio pulsars orbiting each other with a period of only 2.45 hours, led to large improvements in the measurement of relativistic effects, thus validating GR predictions at a level of $1.3\times 10^{-4}$ within $95\%$ confidence level, even with respect to the results from PSR B1913+16. \cite{Kramer2021} The GW emission has been also found in a compact binary system composed  by a high mass neutron star and a very light white dwarf (J0348+0432). The high pulsar mass and the extremely  compact orbit make this system a sensitive laboratory for testing GR and alternative theories. Also in this case, the observed orbital period decay (of only $8\times 10^{-6}$ s yr$^{-1}$) agrees very well with GR predictions, supporting its validity even for the extreme conditions present in the system.Moreover, that  system strengthens recent constraints on the properties of dense matter and provides insight to binary stellar astrophysics and pulsar recycling.\cite{Antoniadis2013}} signaled by the decrease of the orbital period of the system. \cite{Weisberg2016}  The GW170817 event, detected by the LIGO/Virgo interferometers, is  the real pillar of ``multimessenger astronomy'' since it was also detected by a plethora of telescopes, practically in all bands of the electromagnetic spectrum.  \cite{Abbott2017} This discovery, in addition of being extremely important by itself, has allowed to identify the galaxy  where the coalescence occurred and therefore its distance (NGC 4993, at a distance of about 40 Mpc) and to directly confirm that short gamma-ray bursts are the result of the merging process of two neutron stars. It has also allowed to probe a long-standing open problem in astrophysics related to the formation of very heavy elements: supernovae produce large amount of $Fe$ but elements much heavier than $Fe$ cannot form in stellar nucleosynthesis as not enough neutrons are available for the formation of nuclei.  Only very recently it has become clear that the formation of these elements can occur either in the merging process of a pair of neutron stars, as confirmed by the spectroscopic observations in the afterglow of GW170817 which show incontrovertible evidence  that binary neutron star mergers host r-process nucleosynthesis and in the accretion disk surrounding BHs. In this respect we mention that Just et al. \cite{Just2021} have conducted very sofisticated numerical simulations showing that the accretion disks able to ignite the r-processes are only those deriving from two particular astrophysical events: the gravitational collapse of a very massive and rapidly rotating star with a nucleus heavier than about $30\,M_{\odot}$ (called {\it collapsar}) or the merging of two massive neutron stars which gives rise to the {\it kilonova} phenomenon. Both catastrophic events are able to produce a BH surrounded by a dense and hot accretion disk, which is rapidly rotating. The accretion disks where the occurence of the r-processes is more likely are those with mass in the range $0.01-0.1\,M_{\odot}$. 

Until now the attention of the GW Collaborations focused on the event from binary systems. However, especially in the optics of the interferometers planned for the next decade, hyperbolic encounters may play an important role.\cite{Capozziello2008}  The reader is referred to Refs.~\refcite{DeVittori2012,Bini2021} for  detailed analyses of the expected energy spectrum of the emitted GWs in such cases.
The detection of hyperbolic events by GWs, which could be reached also by ground-based intereferometers of the nearby future, \cite{Mukherjee2021}  might shed light on the origin of the components of the emitting systems and constrain, for example, the formation rate of PBHs. \cite{GarciaBellido2018}  

The detection of GW events have allowed estimating the coalescence  rate between black holes and, actually, the strongest constraints on the occurence of such catastrophic events in the  universe derive from GW observations. 
This issue is particularly relevant since it is related to the question about the formation process of the BHs we detect through GW interferometers. Did these BHs formed through stellar evolution in the standard way or  are instead  PBHs? One way to tackle this problem is, first of all, to increase enough the event statistics by detecting an higher number of events. However, one also needs to charatcerize them better in order to constrain the physical  parameters of the coalescing BHs, in particular their orbital eccentricity and the respective spin directions. \cite{Franciolini2021}
In fact, most BHs that are born  in binary systems undergo to an orbital eccentricity decrease due to GW emission, which tends to circularize their orbit. On the other hand,  binary BHs formed through gravitational capture in chance encounters, as mainly expected in the case of PBHs, can form with high initial eccentricity and quite small orbital distance, leaving in most cases insufficient time for the orbital circularization before their coalescing process.\cite{Gayathri2020}
Understanding all these issues is one of the main objectives of the next generation of either ground-based and space-baased  GW interferometers, and it could be a puzzle since multiple formation pathways of BH formation  may occur.\cite{Zevin2021}

One thing to keep in mind is that ground-based laser interferometers, such as LIGO, Virgo and KAGRA (but also the next-coming LIGO-India, planned to start taking data in 2026), are sensitive only to relatively high-frequency GWs, that is to GWs in the frequency range $10-10^4$ Hz. Therefore, these intereferometers can detect only  binary black holes or binary neutron stars at the
final stage of their evolution, just before their coalescence. Third generation ground-based laser interferometers, such as ET
(Einstein Telescope) and CE (Cosmic Explorer), planned to be built in the near future, will be more sensitive than the present day interferometers, but their frequency band will not change sensibly being  in the range $1-10^4$ Hz.
In order to detect compact object systems
in earlier orbital phases, alongside to other types of  GW sources, such as 
extreme mass-ratio inspirals or coalescing massive
black-hole binaries, space-based laser interferometers that bypass the problems due to the Earth seismic noise are needed. The most promising project in this direction is LISA (Laser Interferometer Space Antenna), which will likely consist of a constellation of three satellites, separated by a distance about $2.5 \times 10^6$ km, in a triangular configuration, orbiting around the Sun. Its frequency interval would be in the range $10^{-5}-10^{-1}$ Hz, corresponding to BH binaries orbital periods from roughly a couple of  days to tens of seconds.\cite{AmaroSeoane2017}  

Even if LISA, whose launch is planned in the next decade,
is an extremey advanced detector, it will likely not be able to detect ultra-low frequency GWs in the range  $10^{-10}-10^{-6}$   Hz. These GWs are expected to be generated by many sources of cosmological interest, such as inspiralling SMBHBs, \cite{Rajagopal1995} the inflation phase in the early universe, \cite{Starobinski1979} or even cosmic strings. \cite{Damour2001}
Detecting such GWs is possible, however, through pulsar timing arrays (PTA), which exploit the telescopes generally used for radio astronomy to measure the very tiny variations in the times of arrival (ToA) of the pulses emitted by millisecond pulsars (MSPs), induced by GWs. Indeed, MSPs  are the most precise clocks since the change $\dot{P}$ of their spin period $P$ turns out to be extremely small extremely small ($\dot{P}\simeq 10^{-19}-10^{-20}$ s s$^{-1}$). \cite{Lorimer2008}

A PTA is a set of millisecond pulsars, which are extremely precise and stable clocks, that are constantly monitored by several ground-based radio telescopes in order to collect the pulse ToAs. PTAs are used for many purposes, such as to search for extra-solar planets around pulsars \cite{Wolszczan} or for
ultra-low frequency GWs. \cite{Sazhin} The main PTA collaborations are at present the European Pulsar Timing Array (EPTA), the Indian Pulsar Timing Array (InPTA), the North American Nanohertz Observatory for Gravitational Waves (NANOGrav), and
the Parkes Pulsar Timing Array (PPTA). They join their efforts as the International Pulsar
Timing Array, or IPTA (see Ref.~\refcite{Verbiest2016} and references therein).

These PTAs have timed several tens of millisecond pulsars for more than ten years with an accuracy that should be sufficient, in principle, to detect GWs and some clues of the presence of a common red process compatible with a gravitational wave background has been 
found recently, even if the evidence for such a  background is not  strong enough to claim for a detection. \cite{Arzoumanian2020,Blasi2021,Goncharov2021,Chalumeau2022} The reason
behind that is unclear, as yet, and  could be due to the relatively small number of MSPs
available for PTAs,  to an insufficient observation time spawn,  or  could result from noise that is present in individual pulsars' data not properly modeled in the data analysis. Therefore, it is essential to continue collecting data for many more years. 

Another point to stress is that it is possible that  the standard GW detection technique needs to be complemented by independent methods. 
Indeed, very recently it was proposed, based on the results of numerical simulations, to include millisecond pulsars harbored in the core of some globular clusters in PTAs. The advantage of this proposal is that of taking advantage of the correlated signals amoung these closely packed pulsars, once all the possible time of arrival variations due to the globular cluster have been taken into account in the timing model. This might provide an important step forward towards the GW detection by PTAs. \cite{Maiorano2021a,Maiorano2021b}.

As a final point before closing this Section, it is worth mentioning that lensing effects can appear also for GWs. However, in this case some complication arise since wave optics effects cannot be neglected and simple geometrical optics is not enough to treat the problem correctly. In fact, in this case the wavelength of the GWs is expected to be longer than the Schwarzschild radius of the lens mass. For example, in the case of gravitational  lensing of chirp signals emitted during the coalescence of a supermassive black hole binary at redshift $z\simeq 1$ (just to give an example), which is expected to be in the  frequency band of LISA, wave effects become relevant  in the case of lens masses below about $10^8\,M_{\odot}$ and, if one wants to extract the lens parameters from these kinds of observations, an accurate treatment of the problem is mandatory. \cite{Takahashi2003}
Also in the case of GWs emitted by rotating neutron stars in the Galactic Bulge and lensed  by the BH  Sgr A$^*$ at the Galactic center, diffraction effects may be important. On this issue we refer the reader to Refs. ~\refcite{Ruffa1999,DePaolis2001,DePaolisNucita2002,Congedo2006,Cusin2020}.

\section{Conclusions}

Einstein's GR is one of the towering achievements of physics,  certainly the crowning achievement of classical physics. Even if laboratory experimental tests of Einstein’s theory were slow to come, GR has passed every test (note that only a few of these tests have been discussed  in Sections 2 and 3) physicists have devised so far and, until now, more than a century after  its  pubblication, there are no sign that it does not give always the correct result.\cite{Will2014,Tino2020} All experiments and astronomical observations clearly confirm the triumph of GR in any condition and any strength of the gravitational field, up to the extremely strong fields of the BHs, as the image of the supermassive BH at the center of the M87 galaxy by the EHT collaboration clearly shows (see the discussion in Section 2).

In this respect, the detection of GWs has marked an extremely important step. In fact, the measured signal matches the waveform predictions of Einstein’s GR and allows testing that theory in extreme conditions, not reachable before. So, with GW detection GR has  passed its toughest test.
Unitl nowadays about $10^2$ GW events have been discovered: this tells us not onlt that BHs and binary BHs do exist but also  that they form, evolve and die during a period shorter than the age of the universe. Before the GW detection astronomers had never seen directly binary BHs. Observations also show a dycothomy in the mass of BHs in accreting systems observed in the X-ray band and BHs in binary systems detected through GW interferometers. Why   it is  so  is an important open question.

It is also worth noticing that today the technology behind the detection of GWs, that is that provided by the gravitational wave interferometers, is being used not only to probe inan independent way the expansion of the universe and constrain the cosmological parameters, \cite{GWTC-32021} but also in the long-standing search for dark matter. Thought to make up about $85\%$  of all matter in the universe, dark matter has never been observed directly and remains one of the biggest unsolved mysteries in modern physics. A recently developed method to search for these elusive particles make use of the GW laser interferometers: scalar field dark matter particles would pass  through the Earth causing an almost  impercettible vibration of the mirrors of the interferometers (in particular the  research have been conducted on the UK/German GEO 600 GW detector). The vibrations of the mirrors would disturb the laser beams in a particular way characteristic of the kind of the  dark matter particles. \cite{Vermeulen2021} For the moment these new techniques have only allowed to definitively rule out some kind of dark matter theories, but they certainly  have the  potential to discover dark matter at some point in the near future.

In the previous sections we have described how GR has been tested in different conditions: around the Sun, in the Solar System, around white dwarfs, pulsars and black holes.  GR always passed all the countless tests. It has also been tested on very large scales, that is over distances as large as 1 Gpc. For example, Reyes et al. \cite{Reyes2010} analyzed a survey of about $7\times 10^4$ galaxies  combining for the first time three different measurements: the weak gravitational lensing (thus measuring how much the galaxies' mass was bending light from other galaxies around them), the galaxies speeds, and their clustering properties as a function of the distances. These measurements were combined in order to test GR and other alternative gravity theories, in particular $f(R)$ and TeVeS theories. It goes without saying that GR won the competition, being perfectly consistent with observatiosn of the universe at large scales.

However, in spite of the triumph of GR one can still wonder whether it will someday face the same fate of Newton's theory of gravity. Science, by its nature, never ends, nothing is definitive and absolute, and no physical law is really safe from being called into question.  Even if GR has passed every test so far, nobody knows for sure that GR applies everywhere and in any conditions. Indeed, several rival theories, the so-called alternative thoeries of gravity, have been proposed over the years just in case it does not.

In any case, it is worth studying alternative gravity theories. For example, GR teaches us that inside black holes there exist singularities, where the physical laws break down. However, there  is a hope that a consistent quantum theory of the gravitational field may allow finding a new descriptioin of black holes devoided by singularities. The same holds also for the Big Bang: GR
is a classical theory and should not work well in the quantum regimes, and it cannot therefore be safely applied neither to the singularities inside black holes and at the Big Bang. Quantum gravity would be needed for the purpose, but nobody has been able to consistently quantize GR. Alternative theories might be easier the GR to quantize, but that is only a hope at present.

As a final note, the fact that often Einstein made erroneous claims and  changed his opinion during the years certainly does not make us admire less him.
Probably we admire him even more after that: intelligence, indeed, is not being
stubborn in one's own ideas but to be ready to change them if needed!

\section*{Acknowledgments}
Prof. Asghar Qadir, a mentor for generations of physicists and mathematicians in Pakistan and elsewhere, is warmly acknowledged for the many years of fruitful collaboration and friendness, as well as  for a critical reading of the present manuscript. The INFN projects TAsP and EUCLID are acknowledged for partial support, as well as the organizers of the  Fourth Punjab University International Conference on Gravitation and Cosmology and Prof. Muhammad Sharif in particular.

\end{document}